# Rotational anisotropy Raman spectrometer for high-sensitivity crystallographic symmetry analysis

*Di Cheng, Junxiang Li, Shizhuo Luo, Zehao Chen, and Xinwei Li[*]*

Di Cheng, Junxiang Li, Shizhuo Luo, Zehao Chen, and Xinwei Li
Department of Physics, National University of Singapore, Singapore 117551, Singapore
E-mail: xinweili@nus.edu.sg

Funding: National Research Foundation under award no. NRF-NRFF16-2024-0008

D.C., J.L., and S.L. contributed equally to this work.

**Abstract**

Raman spectroscopy stands as a cornerstone technique for probing collective excitations and emergent quantum phases in solids. While polarization-resolved Raman scattering has been widely used to extract symmetry information of eigenmodes, its conventional geometry suffers from significant limitations: it accesses only a subset of Raman tensor elements, enforces $\pi$-periodic intensity patterns that obscure intrinsic crystalline symmetries, and lacks sensitivity to wavevector-dependent anisotropy. To overcome these constraints, here we introduce rotational-anisotropy Raman spectroscopy (RA-Raman). By measuring scattering intensity during full azimuthal rotation of the optical scattering plane at oblique incidence, this geometry enables complete reconstruction of the Raman tensor and reveals rich rotational anisotropy patterns essential for accessing subtle symmetry information elusive to conventional methods. We developed a prototype instrument to validate this approach experimentally. For centrosymmetric crystals, RA-Raman unambiguously identifies phonon symmetry representations and determines crystallographic axes. In noncentrosymmetric crystals, it resolves directional anisotropy and angular dispersion of phonon-polaritons, enabling quantitative determination of the Faust-Henry coefficient and the separation of deformation-potential and electro-optic scattering contributions. By demonstrating unprecedented symmetry-resolving power in standard benchmark crystals, we establish RA-Raman as a powerful tool with far-reaching potential to discover and characterize symmetry-breaking phases and topological excitations in quantum materials.

# 1. Introduction

Since the discovery of inelastic light scattering (the Raman effect) in 1928, Raman spectroscopy has proven indispensable across disciplines ranging from physical sciences and materials engineering to biological research, environmental monitoring, and food science applications[1–5]. In solid-state systems, the inelastic light scattering process directly resolves fundamental excitations associated with the lattice, charge, and spin degrees of freedom in the frequency domain, unveiling critical material properties such as vibrational modes[1], electronic/magnetic excitations[6–9], carrier density[10], strain profiles[11], and structural phase transitions[12,13].

A defining strength of Raman spectroscopy, particularly effective in single-crystal studies, is its capacity for symmetry analysis. This approach not only facilitates routine tasks such as assigning Raman-active phonon modes and determining crystal point groups but has also driven the discovery of many exotic quantum phases in solids. The recurring framework for these breakthroughs involves identifying characteristic excitations that act as fingerprints of emergent orders: charge-density waves were revealed by order-parameter-coupled phonons[14,15]; the enigmatic hidden order in $URu_2Si_2$ was elucidated through antisymmetric electronic modes[16,17]; superconducting gap symmetry was determined by probing pair-breaking excitations[18]; and ferro-axial order was unveiled via the observation of axial Higgs modes[19]. In each case, symmetry analysis involves determining both the host crystal's symmetry and the transformation properties of the probed eigenmodes, providing decisive evidence for emergent quantum phenomena in condensed matter systems.

Building on the standard procedure of symmetry analysis which is rooted in selection rules and polarization-dependent scattering cross-sections[1,20,21], here we identify inherent limitations in conventional polarization-resolved Raman scattering (P-Raman) measurements — a technique employed in nearly all symmetry-sensitive Raman studies to date. Because only the polarization is rotated at a fixed wavevector, P-Raman measurements probe an incomplete subset of Raman tensor components, enforce $\pi$-periodic angular intensity patterns for all crystal classes, and obscure wavevector-dependent anisotropy.

To overcome these constraints, we introduce a new scheme named rotational-anisotropy Raman spectroscopy (RA-Raman). The setup employs a controlled oblique incidence geometry to measure scattering intensity as a function of optical scattering plane rotation, rather than polarization rotation alone. Guided by a tensor theory analysis, which elucidates the superior symmetry resolution of our design, we developed a prototype instrument and experimentally validated its capabilities using two distinct crystal classes. In centrosymmetric crystals (silicon and sapphire), RA-Raman unambiguously identifies crystal axes and orientations — even on surfaces that appear isotropic to standard polarization-based optical probes, including, most notably, those with threefold rotational symmetry. In noncentrosymmetric crystals (gallium phosphide, α-quartz, and lithium niobate), where the polaritonic effects are active[1], our setup resolves directional anisotropy and dispersion that are coupled with the phonon wavevector to linear order, enabling quantitative extraction of the Faust-Henry coefficient and the angular dispersion relations of phonon-polaritons.

Section 2 is devoted to a description of how RA-Raman differs from standard P-Raman in measurement principles and capabilities, followed by details of our prototype RA-Raman apparatus. Section 3 presents measurements on centrosymmetric and noncentrosymmetric crystals, demonstrating the instrument's distinctive functions. Altogether, by establishing the unprecedented symmetry-resolving power of RA-Raman even on a set of standard, well-understood single crystals, we contend that this method has far-reaching potential to catalyze symmetry-guided scientific discoveries; these implications, together with an outlook, are summarized in Section 4.

## 2. RA-Raman instrument

### 2.1. Instrument design principles

Figure 1a depicts the scattering geometry defined by the RA-Raman setup. An excitation laser beam is focused onto the sample at an oblique incidence angle $\theta$. From the omnidirectional Raman scattering, we selectively collect the Stokes-shifted component of the light that is strictly back-scattered, retracing the path of the incident excitation beam in the opposite direction. Anisotropy contrast is obtained by measuring the Raman intensity as a function of the azimuthal angle $\varphi$, defined as the rotation of the scattering plane (containing the incident beam and its specular reflection) about the sample's surface normal. The incident polarization can be set to linear [s-polarized (S) or p-polarized (P)] or circular [right-circular (R) or left-circular (L)], with the analyzed polarization of the collected light independently configured for the same set of states (S, P, R, L). Geometrically, this configuration is equivalent to rotating the sample at fixed oblique incidence; however, by rotating the optical scattering plane instead, we ensure the laser spot remains stationary on the sample surface — a feature particularly advantageous for small specimens or when the measurement environment precludes physical rotation of the sample.

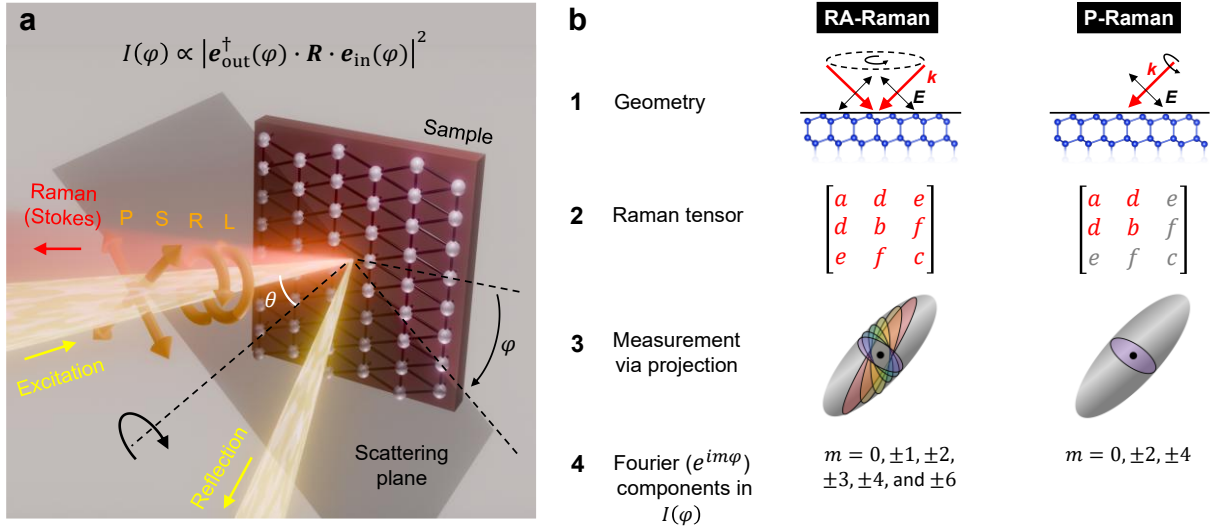

**Figure 1. Conceptual foundation of Rotational-Anisotropy Raman (RA-Raman) spectroscopy. a)** Schematic of the RA-Raman measurement geometry. Scattering intensity anisotropy is measured as a function of the azimuthal angle $\varphi$, while the oblique incidence angle $\theta$ is held constant. **b)** A comparative analysis highlighting the key differences between the RA-Raman and conventional polarization-resolved Raman (P-Raman) schemes: b1) measurement geometry, b2) elements probed in the Raman tensor (red: probed; grey: unprobed), b3) a geometric visualization of the measurement principle, where probing the Raman tensor is analogous to taking 2D sections of a 3D ellipsoid, and b4) the resulting Fourier components in the measured intensity anisotropy.

A crucial feature of the RA-Raman scheme is the use of oblique angles for both the incident light and the analyzed scattered radiation relative to the sample surface. This simple yet previously overlooked arrangement provides markedly greater sensitivity to crystalline symmetry than conventional P-Raman measurements. In the traditional P-Raman configuration, the measurement geometry is identical for polarization states rotated by $\pi$, which forces all polarimetric patterns to exhibit twofold rotational (C$_2$) symmetry, even for crystals that inherently lack it. In contrast, rotating the scattering plane angle $\varphi$ by $\pi$ in the RA-Raman scheme represents a physically distinct and inequivalent measurement geometry (Figure 1b1), thereby breaking the C$_2$ constraint.

The superior symmetry sensitivity of RA-Raman scheme follows directly from the standard Raman tensor theory[22]. The first-order scattering intensity is in general proportional to the squared modulus of the Raman tensor ($\boldsymbol{R}$) contracted with the two polarization vectors of the incident [$\boldsymbol{e}_{\text{in}}(\varphi)$] and scattered fields [$\boldsymbol{e}_{\text{out}}(\varphi)$]:

$$I(\varphi) \propto \left| \boldsymbol{e}_{\text{out}}^{\dagger}(\varphi) \cdot \boldsymbol{R} \cdot \boldsymbol{e}_{\text{in}}(\varphi) \right|^2. \tag{1}$$

The rank-two tensor $R$ encodes the differential change of the optical susceptibility with respect to a displacement along a phonon normal mode, with elements transforming as various irreducible representations of the crystallographic point group. As shown in Figure1b2, the conventional P-Raman technique can access at most 4 of the 9 elements (or 3 of the 6 independent elements if $R$ is symmetric — typical for nonmagnetic, achiral systems in the quasistatic limit). The constraint arises because rotating only the polarization vectors samples tensor components confined to the plane perpendicular to the incident photon wavevector, leaving out-of-plane components inaccessible.

The RA-Raman scheme, in contrast, enables full reconstruction of the $R$ tensor. For each azimuthal angle $\varphi$, the measurement probes 4 elements in coordinate transformed Raman tensor $R' = U^T(\varphi)RU(\varphi)$, where $U(\varphi)$ is the 3D rotation matrix set by the local frame defined by the current wavevector. The $\varphi$ dependence of $U(\varphi)$ ensures that a continuous rotation of the scattering plane systematically mixes all elements of the $R$ tensor into the 4 probed elements in $R'$, allowing all 9 elements of $R$ to be determined through a full rotation. This process can be conceptualized geometrically (Figure1b3). Assuming for visualization that $R$ is symmetric and positive definite (although our conclusion does not rely on this assumption), the tensor can be uniquely represented by a 3D ellipsoid defined by function $x^T R x = 1$, with $x^T = [x \quad y \quad z]$. While P-Raman is limited to observing a single 2D projection of this ellipsoid onto the plane orthogonal to the fixed wavevector — insufficient for determining its three-dimensional form — RA-Raman acquires a series of distinct 2D sections by varying the scattering plane orientation. This comprehensive set of projections fully constrains the ellipsoid's geometry, providing a complete description of the Raman tensor.

In addition to enabling comprehensive interrogation of the Raman tensor, the RA-Raman geometry supports a much richer family of rotational anisotropy patterns in the scattering intensity (Figure 1b4). From Equation (1), the components of the polarization vectors $e_{\text{in}}(\varphi)$ and $e_{\text{out}}(\varphi)$ are functions of the azimuthal angle $\varphi$ that can only be proportional to $\sin \varphi$, $\cos \varphi$, or 1 (i.e. independent of $\varphi$). In a Fourier expansion in $e^{im\varphi}$, these basis functions contribute harmonics with $m = \pm 1$ and 0 for each polarization vector. Consequently, Equation (1) implies that the angle-dependent scattering intensity $I(\varphi)$ — arising from products of these Fourier components contributed by both the incident and analyzed polarizations — can contain harmonics with $m = \pm 4, \pm 3, \pm 2, \pm 1$, and 0, corresponding to patterns compatible with $C_4$, $C_3$, $C_2$, $C_1$, and $C_\infty$ (isotropic) symmetries. While RA-Raman can observe and analyze all of these anisotropy profiles, P-Raman is restricted by its measurement configuration to harmonics $m =$ $\pm 4, \pm 2,$ and 0, because, as mentioned earlier, a $\pi$ rotation in polarization produces an equivalent optical configuration and thus enforces $\pi$-periodicity in the response[21]. Furthermore, in material systems where the inelastic scattering couples to the excitation wavevector, RA-Raman can exhibit $C_6$-symmetric anisotropy, corresponding to $I(\varphi)$ containing harmonics with $m = \pm 6$. This arises when the Raman tensor itself acquires explicit $\varphi$-dependence via wavevector coupling[22],

$$\boldsymbol{R}(\varphi) = \boldsymbol{R}^0 + \overleftrightarrow{\boldsymbol{\chi}} \cdot \boldsymbol{k}(\varphi), \tag{2}$$

where $\boldsymbol{R}^0$ is the $\varphi$-independent part, $\overleftrightarrow{\boldsymbol{\chi}}$ is a third-rank tensor capturing the wavevector dependence with $\chi_{ijk} = \partial R_{ij}/\partial k_k$, and the wavevector $\boldsymbol{k}(\varphi)$ has components proportional to $\sin\varphi$, $\cos\varphi$, or 1, thereby elevating the highest Fourier order in $I(\varphi)$ to $m = \pm 6$. Taken together, the substantially broader set of rotational-anisotropy profiles accessible with RA-Raman, compared to P-Raman, yields markedly enhanced symmetry resolution.

**2.2. Prototype instrument**

We translated the RA-Raman concept into a functional experimental system (Figure 2). A 633-nm excitation laser is directed through a series of optics: a dichroic mirror (DM), a wedged mirror (WM), and flat mirrors (M1, M2), before being focused onto the sample at an oblique incidence angle ($\theta = 10°$) by an objective lens (L3). The key to scanning the scattering plane angle $\varphi$ is the lens-mirror assembly L1-WM-L2. As the WM rotates about the optical axis, it deflects the reflected beam to trace a cone. The matched lens pair L1 and L2 then collimates this beam and converts its conical motion into a pure lateral shift, causing it to trace a cylinder up to L3. We selectively collect the back-scattered Raman signal that retraced the excitation laser path in the opposite direction; this is ensured by placing an aperture (AP) before L3, which transmitted the counter-propagating excitation and back-scattered beams while blocking the intense specular reflection. The scattered beam is then separated from the excitation beam by the DM, collected by fiber optics, and directed to a high-resolution spectrograph for spectral analysis. Polarizations of the incident and analyzed light are selected via linear polarizers (LPs) paired with either a vortex half-waveplate (VWP) or a quarter-waveplate (QWP). The VWP, with its spatially varying fast axis, locks the analyzed linear polarization (S or P) relative to the rotating scattering plane, while the QWP is used for circular polarization channels (R or L). By synchronously rotating the WM and AP about the optical axis, the beams trace a cone between the WM and L2, a cylinder between L2 and L3, and finally a cone between L3 and the sample. This configuration enables full azimuthal rotation of the scattering plane around the sample normal while keeping the laser spot fixed on the sample surface, thereby eliminating the positional drift that would occur if the sample were rotated. Furthermore, to ensure that the oblique incidence angle $\theta$ is precisely maintained while $\varphi$ is scanned, we employ a dedicated alignment procedure: by adjusting the sample's tip-tilt, we match the lateral shift and pointing of the incident excitation and the specularly reflected beams between AP and L3. This alignment ensures that any anisotropy observed in the polar plots faithfully reflects the crystal lattice symmetry and its quantum excitations, free from instrumental artifacts.

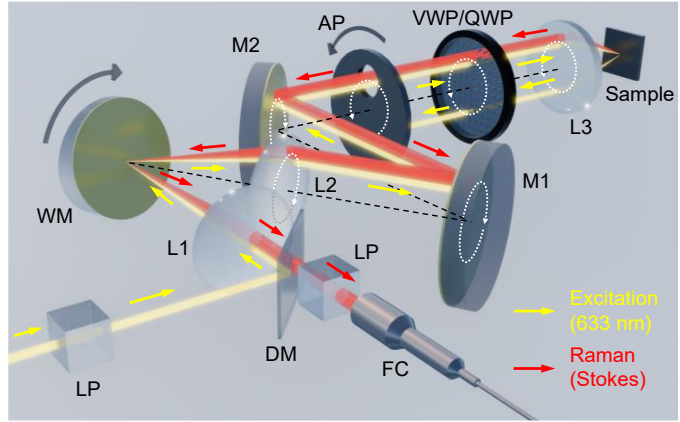

**Figure 2. Schematic of the RA-Raman setup.** WM, wedged mirror; LP, linear polarizer; L1/L2/L3, lenses; M1/M2, mirrors; DM, dichroic mirror; AP, aperture; VWP, vortex half-waveplate (for S and P polarizations); QWP, quarter-waveplate (for R and L polarizations). The collected light is directed to a spectrograph via a fiber coupler (FC).

## 3. Results

### 3.1. Centrosymmetric crystals

To demonstrate the capability of our RA-Raman instrument, we first performed measurements on standard centrosymmetric crystals — sapphire and silicon — and the results are summarized in Figure 3. Centrosymmetric crystals constitute a relatively straightforward class of material systems: first-order Raman-active phonons must be of even parity, and the linear wavevector dependence described by Equation (2) vanishes by inversion symmetry[22]. Accordingly, the standard Raman tensor formalism applies.

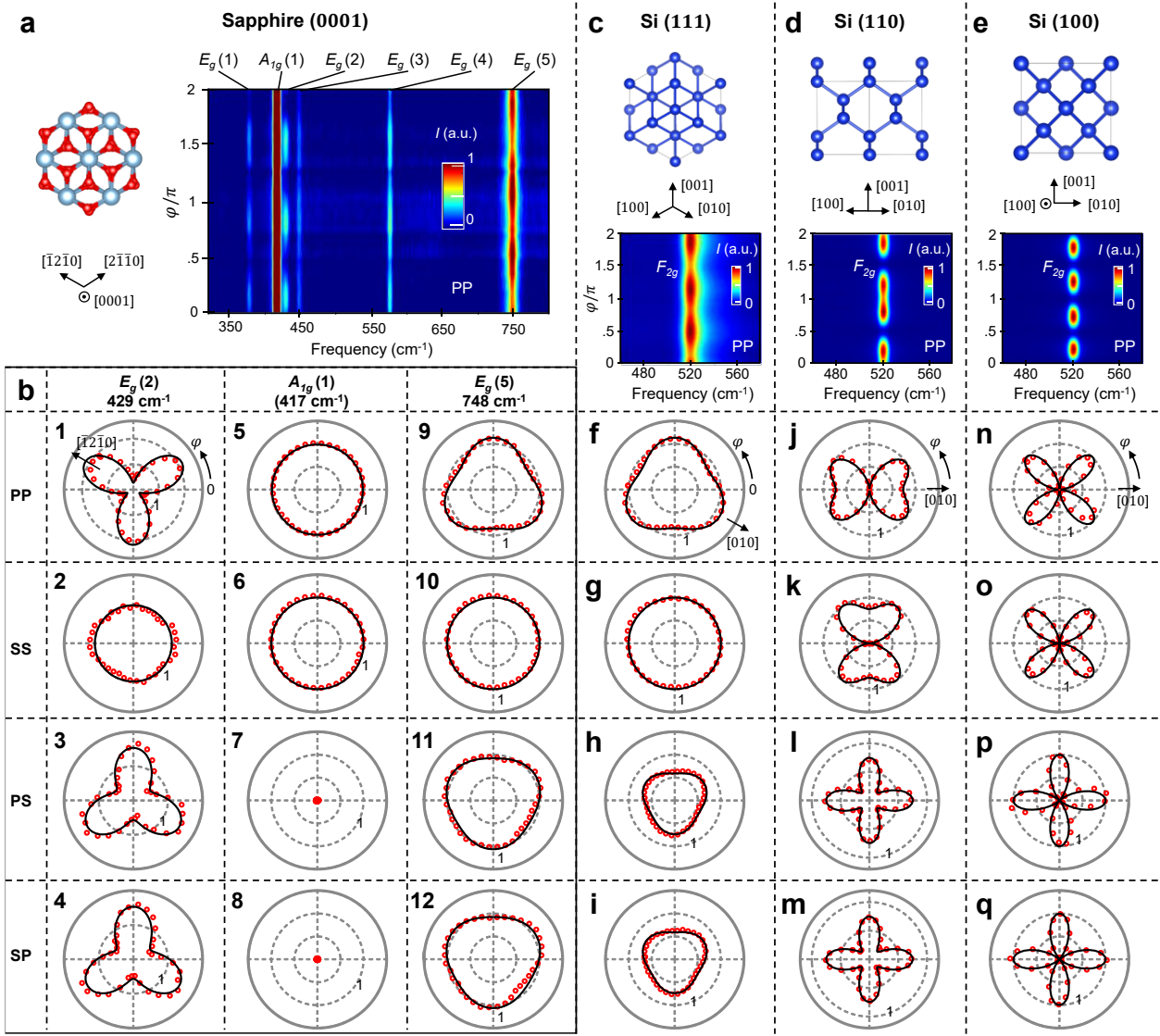

**Figure 3. RA-Raman characterization of centrosymmetric crystals. a)** Atomic structure of the sapphire (0001) surface (left) and the corresponding azimuthal-angle-dependent ($\varphi$) Stokes-scattering spectra in the PP polarization configuration (right). **b)** Polar plots of the $\varphi$-dependent intensity $I(\varphi)$ for the $E_g$(2), $A_{1g}$(1), and $E_g$(5) phonon modes under all polarization configurations (PP, SS, PS, SP). **c-e)** Atomic structures of the **c)** Si (111), **d)** Si (110), and **e)** Si (100) surfaces. **f-i), j-m), n-q)**, Corresponding polar plots of $I(\varphi)$ for the $F_{2g}$ phonon mode across all polarization channels for the Si (111), Si (110), and Si (100) cuts, respectively. Polarization labels for **(f)-(q)** correspond to those shown in **(b)**.

Our measurements on a sapphire (Figures 3a and 3b) demonstrate RA-Raman's ability to fully resolve phonon-mode symmetry. Sapphire ($\alpha$-Al$_2$O$_3$) crystallizes in the trigonal corundum structure with point group $D_{3d}$, in which Raman-active phonons transform as $A_{1g}$ and $E_g$ irreducible representations (irreps). The sample was *c*-cut, i.e. sapphire (0001), with two in-plane crystallographic directions along [$\bar{1}2\bar{1}0$] and [$2\bar{1}\bar{1}0$], and the out-of-plane [0001] direction coinciding with the threefold rotational axis (C$_3$). The intensity map in Figure 3a shows $\varphi$-dependent Raman spectra in the PP measurement configuration (incident: P, detection: P). A total of five $E_g$ modes and one $A_{1g}$ mode are observed, consistent with prior reports[23].

Notably, azimuthal-angle dependence of scattering intensity $I(\varphi)$ differs among modes. To quantify this, we extracted peak spectral weights by fitting each spectrum to a sum of multiple Lorentzian peaks over a broad background. Figure 3b summarizes $I(\varphi)$ for three representative modes — the $E_g(2)$ mode at 429 cm$^{-1}$, the $A_{1g}(1)$ mode at 417 cm$^{-1}$, and the $E_g(5)$ mode at 748 cm$^{-1}$ — under four polarization configurations: PP, SS (incident: S, detection: S), PS (incident: P, detection: S), and SP (incident: S, detection: P). Both $E_g$ modes (and, more generally, all $E_g$ modes) exhibit C$_3$-symmetric patterns in the PP, PS and SP configurations, whereas the SS channel is isotropic. This behavior is captured by Equation (1), using polarization vectors appropriate to the RA-Raman geometry, and by modelling the signal as the incoherent sum of contributions from the two Raman tensors of the doubly degenerate $E_g$ mode

$$\boldsymbol{R}^{Eg,1} = \begin{bmatrix} c & 0 & 0 \\ 0 & -c & d \\ 0 & d & 0 \end{bmatrix}, \boldsymbol{R}^{Eg,2} = \begin{bmatrix} 0 & -c & -d \\ -c & 0 & 0 \\ -d & 0 & 0 \end{bmatrix}. \tag{3}$$

The tensor elements $c$ and $d$ are adjusted to fit the experimental polar plots of $I(\varphi)$ (red circles) across all polarization configurations simultaneously (black lines). By contrast, the $A_{1g}$ mode shows isotropic response in the PP and SS channels (Figures 3b5 and 3b6) and negligible intensity in the PS and SP channels (Figures 3b7 and 3b8), consistent with the diagonal Raman tensor

$$\boldsymbol{R}^{A_{1g}} = \begin{bmatrix} a & 0 & 0 \\ 0 & a & 0 \\ 0 & 0 & b \end{bmatrix}, \tag{4}$$

which forbids cross-polarized scattering. Notably, the robust C$_3$-symmetric polar patterns observed for $E_g$ modes — despite differences in peak-to-node ratios among them — faithfully reflect the trigonal symmetry of sapphire (see surface atomic arrangement in Figure 3a) and can be leveraged to determine the three equivalent in-plane axes. Such symmetry resolution is inaccessible to standard P-Raman, since $I(\varphi)$ in that geometry must not contain the $m = \pm 3$ harmonics as mentioned in Section 2.

To further demonstrate RA-Raman's capability to resolve crystal orientation and identify in-plane axis, we conducted measurements on silicon crystals cut along (111), (110), and (100). Silicon has the diamond cubic structure ($O_h$ point group), with C$_3$ rotation axes along the body diagonals $\langle 111 \rangle$, C$_2$ rotation axes along $\langle 110 \rangle$, and C$_4$ axes along $\langle 100 \rangle$; these surface symmetries are illustrated in Figures 3c-e. The $\varphi$-dependent intensity maps in PP configuration for all three cuts are dominated by the zone-center $F_{2g}$ phonon at 520 cm$^{-1}$, yet the dependence of its intensity on $\varphi$ differs distinctly among the cuts. The $I(\varphi)$ in all polarization channels, extracted via Lorentzian fits, are plotted in Figures 3f–i for Si (111), Figures 3j–m for Si (110), and Figures 3n–q for Si (100).

As a general rule, the common symmetry manifested by the $I(\varphi)$ polar plots across all four polarization channels reflects the crystal's surface symmetry: Si (111) preserves $C_3$ symmetry, Si (110) preserves $C_2$ symmetry, and Si (100) preserves $C_4$ symmetry, in exact accord with the atomic arrangements of the respective surfaces. Moreover, the orientations of the strong lobes guide identification of the in-plane axes. For Si (111), the three equivalent in-plane directions — given by the projections of [001], [100], and [010] onto the (111) surface — align with the strong lobes in PP. This characteristic $C_3$-symmetric pattern arises specifically from the RA-Raman geometry and follows from computing $I(\varphi)$ using the sum of contributions from the three Raman tensors of the triply degenerate $F_{2g}$ mode

$$R^{F_{2g},1} = \begin{bmatrix} 0 & 0 & 0 \\ 0 & 0 & d \\ 0 & d & 0 \end{bmatrix}, \quad R^{F_{2g},2} = \begin{bmatrix} 0 & 0 & d \\ 0 & 0 & 0 \\ d & 0 & 0 \end{bmatrix}, \quad R^{F_{2g},3} = \begin{bmatrix} 0 & d & 0 \\ d & 0 & 0 \\ 0 & 0 & 0 \end{bmatrix}. \tag{5}$$

Analogous calculations for the other cuts show that the strong lobes in the SS channel align with [001] in Si (110), and the strong lobes in the PS and SP channels align with the [001] and [010] axes in Si (100). Together, these measurements on silicon demonstrate that the RA-Raman geometry provides enhanced capability to resolve crystal cut and orientation compared with conventional approaches.

### 3.2. Noncentrosymmetric crystals

Upon application to noncentrosymmetric crystals, the RA-Raman scheme readily reveals a broader range of phenomena. The key distinction from centrosymmetric systems is that, in the absence of inversion symmetry, the mutual-exclusion rule no longer applies: zone-center phonons can be both Raman- and infrared-active. Their atomic displacements induce macroscopic dipole moments and, hence, long-range internal electric fields whose amplitude and polarization depend on the phonon eigenvector. These fields couple the lattice vibrations to the propagating electromagnetic mode in the crystal, so that the Raman-active modes manifest phonon-polariton physics[1,24].

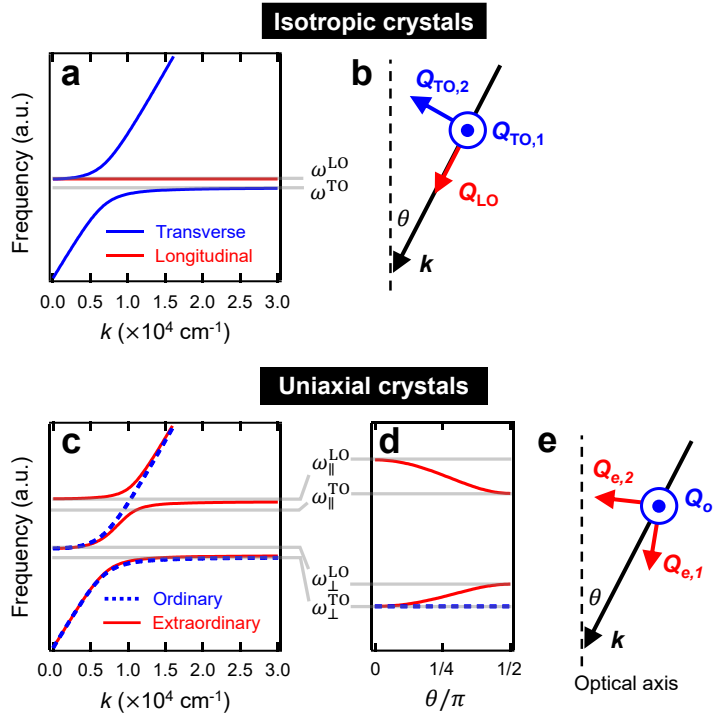

**Figure 4. Theoretical framework for phonon-polaritons in noncentrosymmetric crystals. a)** Dispersion relation for an isotropic crystal, showing the transverse-optical (TO, blue) and longitudinal-optical (LO, red) branches. The frequencies $\omega^{TO}$ and $\omega^{LO}$ represent the asymptotic limits at large wavevector $k$. **b)** Schematic of the phonon polarization vectors for the TO ($Q_{TO,1/2}$) and LO ($Q_{LO}$) modes in an isotropic crystal. **c)** Dispersion relations for a uniaxial crystal, showing the ordinary (blue) and extraordinary (red) polariton branches and **d)** the angular dispersion (at large-$k$ limit) of the polariton branches. **e)** Phonon polarization vectors for the ordinary ($Q_o$) and extraordinary ($Q_{e,1/2}$) modes in a uniaxial crystal. In **(c)-(e)**, the wavevector $k$ forms an angle $\theta$ with the optical axis. The asymptotic frequencies $\omega_\perp^{TO}$, $\omega_\perp^{LO}$, $\omega_\parallel^{TO}$, and $\omega_\parallel^{LO}$ denote the TO and LO limits for polarizations perpendicular ($\perp$) and parallel ($\parallel$) to the optical axis.

Figures 4a and 4b summarize the standard picture of phonon-polaritons in isotropic media. For a given phonon wavevector $k$, a zone-center mode splits according to the orientation of its atomic polarization relative to $k$. The branch with polarization ($Q_{TO}$) transverse to $k$ — the transverse-optical (TO) mode — hybridizes with the photon and exhibits an avoided crossing near the light line $\omega = ck/n_{op}$ ($c$: speed of light, $n_{op}$: optical refractive index), producing two polariton branches. By contrast, the longitudinal-optical (LO) mode, with polarization ($Q_{LO}$) parallel to $k$, does not couple to light and is shifted in frequency relative to the TO mode by the macroscopic dipole-dipole interaction (the LO-TO splitting). In a backscattering geometry, the phonon wavevector magnitude is set by the optical wavevectors of the incoming and outgoing photons, giving $k = 4\pi n_{op}/\lambda$, where $\lambda$ is the probe wavelength; this represents the large-$k$ regime, in which the lower TO polariton branch asymptotically approaches frequency $\omega^{TO}$, while the LO branch remains pinned at $\omega^{LO}$. RA-Raman is ideally suited to interrogate polariton physics because, from a symmetry standpoint, polaritons stem from the explicit wavevector-coupling term in Equation (2); by sweeping the azimuth $\varphi$ of the scattering plane, RA-Raman varies the orientation of $\mathbf{k}$ and thereby resolves the resultant wavevector-dependent directional anisotropy.

Below, we first outline the theoretical framework of Raman scattering from phonon-polaritons in noncentrosymmetric uniaxial crystals (Figure 4c-e), which encompasses our samples; isotropic noncentrosymmetric crystals emerge as a special case of this model. We introduce this theory to identify polariton phenomena that are difficult to probe by conventional methods, and then turn to our experiments to highlight the unique insights provided by RA-Raman.

In a general geometry where the phonon wavevector $\mathbf{k}$ makes an angle $\theta$ with the optical axis of the uniaxial crystal (Figure 4e). The coupled lattice-electromagnetic equations governing the phonon-polariton eigenmodes are[25,26]

$$-\omega^2 Q_\| + \left(\omega_\|^{\text{TO}}\right)^2 Q_\| = a_\| E_\| \tag{6a}$$

$$-\omega^2 Q_\perp + \left(\omega_\perp^{\text{TO}}\right)^2 Q_\perp = a_\perp E_\perp \tag{6b}$$

$$P_\| = a_\| Q_\| + \varepsilon_0(\varepsilon_\|^\infty - 1)E_\| \tag{6c}$$

$$P_\perp = a_\perp Q_\perp + \varepsilon_0(\varepsilon_\perp^\infty - 1)E_\perp \tag{6d}$$

$$\mathbf{E} = \frac{-\mathbf{k}(\mathbf{k} \cdot \mathbf{P}) + (\omega^2/c^2)\mathbf{P}}{\varepsilon_0\left(k^2 - \frac{\omega^2}{c^2}\right)} \tag{6e}$$

Here, $Q$, $E$, $P$, $a$, $\omega^{\text{TO}}$, $\varepsilon^\infty$ denote, respectively, the phonon polarization, electric field, macroscopic polarization, phonon effective charge, TO mode frequency, and high-frequency dielectric constant; subscripts "∥" and "⊥" indicate components parallel and perpendicular to the optical axis. Equations (6a) and (6b) follow from the lattice equations of motion; Equations (6c) and (6d) express the two contributions to the macroscopic polarization — lattice displacement and field-induced electronic polarization; Equation (6e) is obtained from Maxwell's equations in Fourier space.

Prior work[25] has established two solution classes to Equation (6). The first, ordinary polaritons, features lattice polarization $\mathbf{Q}_o$ perpendicular to both the optical axis and $\mathbf{k}$; because $\mathbf{Q}_o$ lies in the isotropic plane, the dispersion is angle-independent and reduces to

$$\varepsilon_\perp^\infty \omega^4 - \omega^2 c^2 k^2 - \varepsilon_\perp^\infty \omega^2 (\omega_\perp^{LO})^2 + c^2 k^2 (\omega_\perp^{TO})^2 = 0, \tag{7}$$

where $\omega_\perp^{LO}$ denotes the frequency of the LO mode polarized perpendicular to the optical axis, and the branches exhibit an avoided crossing (blue dashed curves in Figure 4c), with the lower branch asymptotically approaching $\omega_\perp^{TO}$.

The second class comprises extraordinary polaritons, whose lattice polarization $\boldsymbol{Q_e}$ lies in the plane spanned by the optical axis and $\boldsymbol{k}$; for $\theta \neq 0$, $\boldsymbol{Q_e}$ is generally not perpendicular to $\boldsymbol{k}$, so the mode carries mixed transverse and longitudinal components in its eigenvector. These modes are obtained by recasting Equation (6) into the linear system of equations[26]

$$\begin{pmatrix} (\omega_\parallel^{TO})^2 - \omega^2 + \frac{b}{\varepsilon_0}(\cos\theta\, a_\parallel)^2 & \frac{b}{\varepsilon_0}\cos\theta\sin\theta\, a_\parallel a_\perp & -\varepsilon_\perp^\infty b \sin\theta\, a_\parallel \\ \frac{b}{\varepsilon_0}\cos\theta\sin\theta\, a_\parallel a_\perp & (\omega_\perp^{TO})^2 - \omega^2 + \frac{b}{\varepsilon_0}(\sin\theta\, a_\perp)^2 & \varepsilon_\parallel^\infty b \cos\theta\, a_\perp \\ -\frac{1}{\varepsilon_0 \varepsilon_\parallel^\infty}\omega^2 \sin\theta\, a_\parallel & \frac{1}{\varepsilon_0 \varepsilon_\perp^\infty}\omega^2 \cos\theta\, a_\perp & -\omega^2 + \frac{k^2 c^2}{\varepsilon_\parallel^\infty \varepsilon_\perp^\infty b} \end{pmatrix} \begin{pmatrix} Q_\parallel \\ Q_\perp \\ E_T \end{pmatrix} = 0 \tag{8}$$

with $b = [\varepsilon_\parallel^\infty (\cos\theta)^2 + \varepsilon_\perp^\infty (\sin\theta)^2]^{-1}$, and $E_T = E_\parallel \sin\theta - E_\perp \cos\theta$ is the transverse component of the electric field. Casting Equation (8) as an eigenvalue problem yields the extraordinary polariton eigenfrequencies — directly related to the observed Raman peak positions — as well as the corresponding eigenvectors,

$$\begin{pmatrix} Q_{e,1} \\ Q_{e,2} \\ Q_{e,3} \end{pmatrix} = \boldsymbol{U}^{-1} \begin{pmatrix} Q_\parallel \\ Q_\perp \\ E_T \end{pmatrix} = \boldsymbol{U}^{-1} \begin{pmatrix} Q_z \\ Q_x \cos\varphi + Q_y \sin\varphi \\ E_T \end{pmatrix} \tag{9}$$

where $\boldsymbol{U}$ is the diagonalizing matrix[26]. These eigenvectors determine how the Raman tensor is constructed from basis modes ($Q_{x,y,z}$), which is crucial for analysis of the angle-dependent scattering intensity.

Two notable features emerge naturally for extraordinary polaritons, both difficult to access with conventional Raman spectroscopy. First, the Raman tensor associated with an extraordinary polariton contains two distinct contributions — a deformation-potential term and an electro-optic term[26–29] — so that for mode index $j$,

$$\boldsymbol{R_{e,j}} = \boldsymbol{R_{j,\text{deformation}}} + \boldsymbol{R_{j,\text{EO}}} \tag{10}$$

$$R^{\sigma\rho}_{j,\text{deformation}} = \sum_{m=x,y,z} \frac{\partial \chi_{\sigma\rho}}{\partial Q_m} \frac{\partial Q_m}{\partial Q_{e,j}}$$

$$R^{\sigma\rho}_{j,\text{EO}} = -\frac{b}{\varepsilon_0}\left(\cos\theta\ a_\parallel \frac{\partial Q_\parallel}{\partial Q_{e,j}} + \sin\theta\ a_\perp \frac{\partial Q_\perp}{\partial Q_{e,j}}\right) \sum_{\gamma=x,y,z} \chi^{EO}_{\sigma\rho\gamma} k_\gamma$$

Here, $\chi_{\sigma\rho}$ is the linear susceptibility tensor, and $\partial\chi_{\sigma\rho}/\partial Q_m$ are the Raman tensors of tabulated zone-center modes; the coefficients $\partial Q_m/\partial Q_{e,j}$ and $\partial Q_{\parallel/\perp}/\partial Q_{e,j}$ follow from the eigenvectors derived from Equation (9). The deformation-potential term projects the polariton's atomic polarization onto the standard, tabulated zone-center phonon modes, making explicit how each basis mode contributes to the observed scattering. The second term reflects the electro-optic contribution: infrared activity endows the polariton with a longitudinal field component parallel to the wavevector, which mixes with the driving optical field through the third-rank electro-optic tensor $\chi^{EO}_{\sigma\rho\gamma}$ and appears in Raman spectra as a purely optical effect. Fully determining $\boldsymbol{R}_{e,j}$ — and disentangling displacement-potential from electro-optic contributions — has been challenging.

The second notable feature is the angular dispersion of extraordinary polaritons. Figure 4c shows the full polariton dispersion at a generic incidence angle $\theta$. Unlike ordinary modes, extraordinary polariton frequencies depend on $\theta$ because the excitation partitions between anisotropic transverse and longitudinal components in a $\theta$-dependent manner. As illustrated in Figure 4d, in the large-$k$ limit the lower extraordinary branch evolves from $\omega_\perp^{\text{TO}}$ toward $\omega_\perp^{\text{LO}}$, while the upper branch evolves from $\omega_\parallel^{\text{LO}}$ toward $\omega_\parallel^{\text{TO}}$ as $\theta$ varies from 0 to $\pi/2$. Probing this behavior requires varying the wavevector orientation during Raman acquisition — a capability not implemented in conventional instruments.

Understanding these challenges, we now describe how RA-Raman measurements directly address them. Our samples comprise gallium phosphide (GaP), α-quartz, and lithium niobite (LiNbO$_3$). While gallium phosphide is isotropic, α-quartz and LiNbO$_3$ belong to the uniaxial crystal class with broken inversion symmetry. Owing to the abundance of phonon peaks in these materials, RA-Raman produces extensive datasets; accordingly, we focus on — and analyze in depth — only those modes that unambiguously manifest phonon-polariton physics.

Figures 5a and 5b present RA-Raman measurements on a (110)-cut GaP crystal. GaP crystallizes in the cubic zinc blende structure with tetrahedral point group $T_d$, and its first-order Raman lines are sharp and well documented[30]; GaP also featured in the earliest Raman study of phonon-polaritons[31]. Figure 5a shows the $\varphi$-dependent RA-Raman spectra in the SS configuration, highlighting the transverse-optical (TO, 365 cm$^{-1}$) and longitudinal-optical (LO, 402 cm$^{-1}$) modes associated with the $F_2$ phonon at the zone center. Unlike prior near-forward scattering experiments that mapped polariton dispersion near the light line, our backscattering geometry accesses the large-$k$ limit of Figure 4a, where the TO and LO frequencies approach $\omega^{TO}$ and $\omega^{LO}$, respectively.

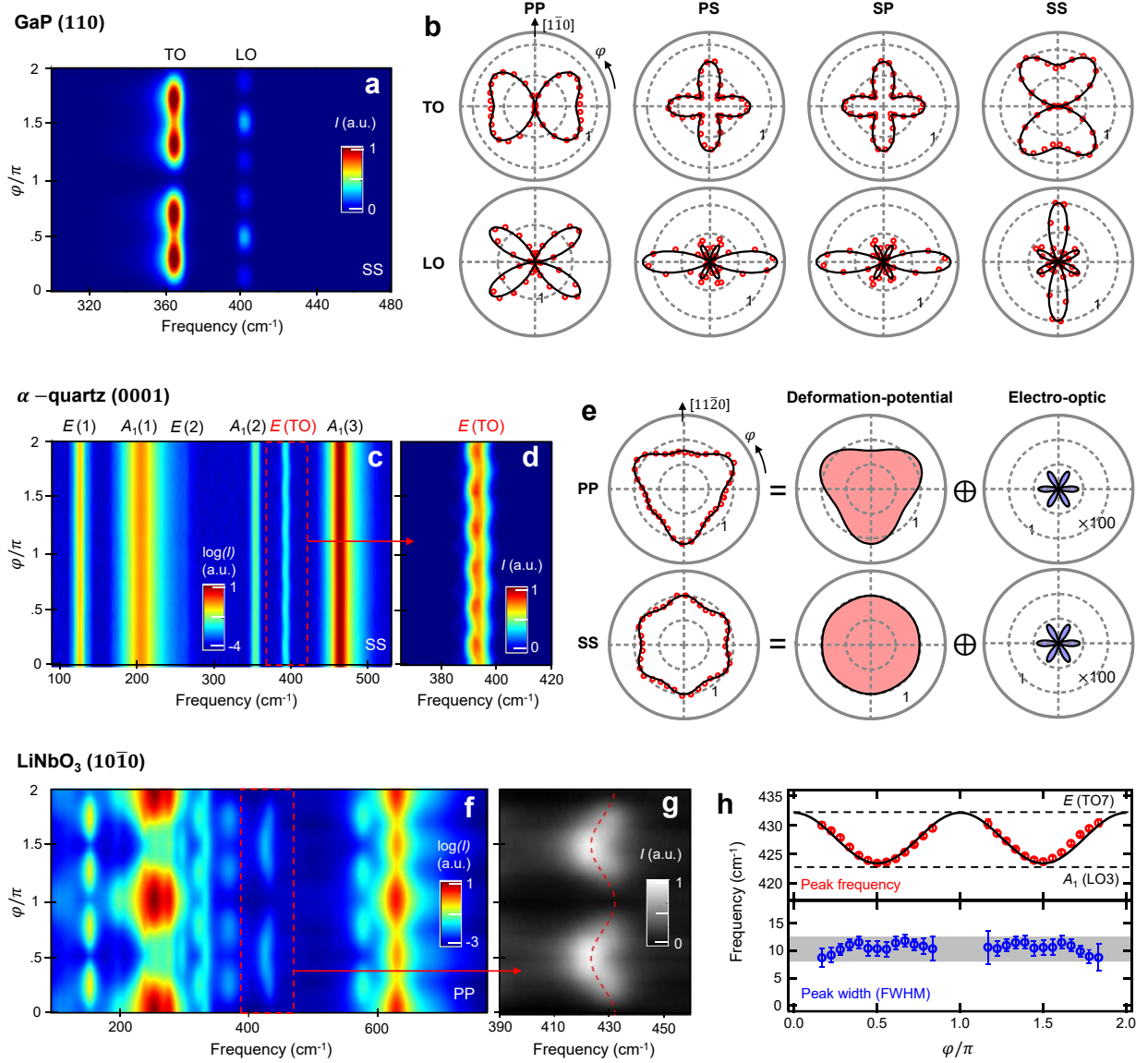

**Figure 5. RA-Raman characterization of phonon-polaritons in noncentrosymmetric crystals a)** $\varphi$-dependent spectra in the SS channel for a GaP (110) crystal. **b)** Polar plots of $I(\varphi)$ for GaP's TO and LO modes in all polarization configurations. **c)** $\varphi$-dependent spectra in the SS channel for a $\alpha$-quartz (0001) crystal. **d)** Zoom-in of the spectral window around the $E$ (TO) mode from **(c)**. **e)** Polar plots of $I(\varphi)$ for the $E$ (TO) mode, showing experimental data (PP and SS channels) and theoretical fits. The response is modeled as a coherent sum of deformation-potential (red) and electro-optic (blue) contributions, which interfere destructively. **f)** $\varphi$-dependent spectra in the PP channel for a $y$-cut LiNbO$_3$ crystal. **g)** Zoom-in of the frequency window around 430 cm$^{-1}$ from **(f)**, showing a mode with strong angular dispersion. **h)** The peak frequency (upper panel) and full width at half maximum (FWHM, lower panel) of the dispersing mode in **(g)** as a function of $\varphi$. In **(b)**, **(e)**, and **(h)**, circled markers are experimental data and solid black lines are theoretical fits. In **(c)** and **(f)**, the logarithm of the intensity is shown to enhance the visibility of weak modes.

The added insight afforded by RA-Raman is evident in the mode-intensity polar plots. Figure 5b displays $I(\varphi)$ for the TO and LO modes under PP, PS, SP, and SS configurations. The TO polar patterns exhibit the expected twofold ($C_2$) symmetry, consistent with the atomic arrangement on the (110) surface. By contrast, the LO mode shows markedly richer anisotropy: while still respecting $C_2$ symmetry, the PS, SP, and SS channels display six nodes and six petals — an anisotropy not reported in prior Raman studies — indicating the presence of $m = \pm 6$ harmonics in the Fourier expansion of $I(\varphi)$. As discussed in Section 2, such components arise only when the Raman tensor is linearly coupled to the wavevector [Equation (2)] — a symmetry hallmark of polariton modes in noncentrosymmetric crystals[32]. By rotating the scattering plane — and thereby the orientations of the incident and scattered optical wavevectors — our RA-Raman instrument varies the phonon wavevector and directly captures this six-petal structure, establishing the wavevector-dependent directional anisotropy of the scattering.

Measurements of the directional anisotropy in $I(\varphi)$ enable quantitative separation of the deformation-potential and electro-optic contributions to inelastic light scattering by polariton modes. We demonstrate this by using $\alpha$-quartz, a uniaxial noncentrosymmetric crystal with trigonal point group $D_3$. A $c$-cut sample [(0001) orientation] was used, so the angle between the phonon wavevector $\boldsymbol{k}$ and the optical axis remains fixed at the incidence angle $\theta$, while $\varphi$ is swept to capture directional anisotropy. From the full set of $\varphi$-dependent RA-Raman spectra acquired under all polarization configurations, Figure 5c shows the SS channel, where three $A_1$ modes [$A_1$ (1) at 207 cm$^{-1}$, $A_1$ (2) at 355 cm$^{-1}$, and $A_1$ (3) at 465 cm$^{-1}$] and three $E$ modes [$E$ (1) at 128 cm$^{-1}$, $E$ (2) at 263 cm$^{-1}$, and $E$ (TO) at 394 cm$^{-1}$] are observed within the measured frequency range[33,34].

Within the $D_3$ point group, $A_1$ modes are Raman-active only, whereas $E$ modes are both Raman- and infrared-active and can therefore exhibit phonon-polariton behavior. Prior measurements show that the $E$ (1) and $E$ (2) modes have insufficient oscillator strength to produce observable LO-TO splitting[33,34]; consequently, their interpretation — like that of all $A_1$ modes — follows the standard Raman-tensor formalism applicable to centrosymmetric systems[35]. By contrast, the $E$ (TO) mode (394 cm$^{-1}$) has a corresponding LO resonance at 403 cm$^{-1}$ and does display polaritonic effects, so it is the focus of our analysis. An enlarged view of this mode (Figure 5d) reveals a clear sixfold modulation of intensity as $\varphi$ is swept through $2\pi$, strongly indicating wavevector-dependent directional anisotropy of the kind identified in GaP.

To quantitatively determine the deformation-potential and electro-optic contributions to the $E$ (TO) mode, we solved the linear equation set Equation (8) using experimentally determined pararmeters: $\omega_\perp^{TO} = 393$ cm$^{-1}$, $\omega_\perp^{LO} = 403$ cm-1, $\omega_\parallel^{TO} = 356$ cm-1, $\omega_\parallel^{LO} = 376$ cm-1, $\varepsilon_\perp^\infty = 2.5$, $\varepsilon_\parallel^\infty = 2$. The effective charge was calculated as $a_{\perp/\parallel} = \sqrt{\varepsilon_0 \varepsilon_{\perp/\parallel}^\infty [(\omega_{\perp/\parallel}^{LO})^2 - (\omega_{\perp/\parallel}^{TO})^2]}$. $\omega_\parallel^{TO}$ and $\omega_\parallel^{LO}$ were taken as the TO and LO frequencies of the $A_2$ mode observed in infrared absorption measurements[34]. The eigenvectors derived in Equation (9) fix the tensor structure of the deformation-potential Raman term $R_{j,\text{deformation}}^{\sigma\rho}$, which reduces to only two adjustable tensor elements. As shown in Figure 5e, this mechanism alone cannot account for the measured $I(\varphi)$ polar plots: clear discrepancies remain in the PP and SS configurations, and the calculated patterns reproduce only the slowly varying envelope but not the fine features at harmonic $m = \pm 6$. These high-order modulations require inclusion of the electro-optic contribution.

The dataset can be fully explained only by including an electro-optic contribution that superposes with — and interferes with — the deformation-potential term. This electro-optic channel is active for $E$ (TO) under our oblique-incidence geometry, which mixes a finite LO component into a mode that is otherwise predominantly TO ($Q_{e,2}$ mode in Figure 4e). To capture this effect, we solved for the electro-optic Raman tensor $R_{j,\text{EO}}^{\sigma\rho}$ and, crucially, derived the symmetry-compatible electro-optic tensor $\chi_{\sigma\rho\gamma}^{EO}$; with $D_3$ symmetry, only four components are nonzero — $\chi_{yyx}^{EO}$, $\chi_{yxz}^{EO}$, $\chi_{yzx}^{EO}$, $\chi_{zyx}^{EO}$ — with all others vanishing[35]. With physically consistent parameter choices for $\chi_{\sigma\rho\gamma}^{EO}$, the electro-optic term supplies the fast six-fold modulation observed in the polar patterns. A coherent sum of the deformation-potential tensor $R_{j,\text{deformation}}^{\sigma\rho}$ and the electro-optic tensor $R_{j,\text{EO}}^{\sigma\rho}$ reproduces the data across all configurations (Figure 5e; see Supporting Information[35] for full datasets and fits).

The task of disentangling the deformation-potential and electro-optic contributions to polariton Raman scattering is known to be important yet nontrivial. The ratio between the two contributions is quantified by the Faust-Henry coefficient[36], a material parameter that bridges Raman and infrared dielectric response, with implications for evaluating carrier concentration, carrier mobility, and nonlinear optical properties of semiconductors[26–29]. Traditionally, this coefficient is inferred from LO-to-TO intensity ratios measured in a fixed geometry, together with estimates of optical transmission/reflection and other setup-dependent factors[26–29] — assumptions that are highly configuration-sensitive and difficult to reproduce. With RA-Raman, the full angular and polarization dependence of $I(\varphi)$ enables us to determine, within a single dataset, the separate deformation-potential and electro-optic contributions and thus extract the Faust-Henry coefficient. This capability opens up a realm of polariton studies that would benefit substantially from the RA-Raman methodology.

Finally, we demonstrate RA-Raman's capability to map the angular dispersion of polariton modes using a *y*-cut LiNbO$_3$ crystal. LiNbO$_3$ is a polar, uniaxial material in the trigonal point group $C_{3v}$. For a *y*-cut specimen with surface normal $(10\bar{1}0)$, the angle between the phonon wavevector ***k*** and the optical *c*-axis $(0001)$ varies as the azimuth $\varphi$ of the scattering plane is swept in our oblique-incidence geometry. This configuration enables direct interrogation of TO-LO mixing and the angular frequency dispersion effect depicted in Figure 4d.

Figure 5f shows the $\varphi$-dependent RA-Raman spectra in the PP configuration, where a number of modes with distinct rotational-anisotropy patterns are readily discernible by eye. Despite the well-known complexity of LiNbO$_3$ — hosting at least nine TO and nine LO modes with substantial LO-TO splittings[37] — we focus on the spectral window near 430 cm$^{-1}$ (Figure 5g), where both the mode intensity and, crucially, the mode frequency exhibit pronounced $\varphi$ dependence. This region coincides with literature assignments of $A_1$ (LO3) and $E$ (TO7) at 422 cm$^{-1}$ and 432 cm$^{-1}$, respectively[37]. As shown in Figure 5h (upper panel) and the red dashed line in Figure 5g, our RA-Raman data show that the observed mode disperses between these two frequencies with an angular variation well described by the empirical relation

$$\omega^2 = (\omega^{\text{TO}} \cos \varphi)^2 + (\omega^{\text{LO}} \sin \varphi)^2 \tag{11}$$

consistent with the standard theory describing TO-LO mixing as the polariton wavevector varies[1]. We further rule out a trivial scenario where our observation results from two modes with fixed frequencies exchanging spectral weight. Linewidth analysis reveals no systematic $\varphi$ dependence of the full width at half maximum (Figure 5h, lower panel), indicating that the observed mode corresponds to a single mode of mixed TO-LO character rather than a superposition of two independent peaks. The angular dispersion effect is therefore demonstrated.

## 4. Conclusion and Discussion

We have introduced a RA-Raman measurement scheme and demonstrated its symmetry-resolving power on a set of standard, well-understood single crystals, readily revealing physical effects that are difficult to probe via traditional means. In centrosymmetric crystals, the technique unambiguously identifies phonon irreducible representations, crystal orientations, and in-plane axes. In noncentrosymmetric crystals, it resolves directional anisotropy and dispersion of phonon-polariton modes arising from the Raman tensor's linear coupling to the phonon wavevector, enabling quantitative determination of the Faust-Henry coefficient and polariton angular dispersion relations.

Looking ahead to broader applications, we believe this method has far-reaching potential in the following realms of condensed-matter physics:

(1) Symmetry-breaking phases of matter. RA-Raman's symmetry-resolving power can accelerate the discovery and diagnosis of symmetry-breaking quantum phases of solids. Order parameters associated with quantum magnetism[7,8], charge-density waves[14,15], superconductivity[18], and multipolar orders[38] necessarily reduce a parent point-group symmetry; RA-Raman can detect these changes while operating under the extreme conditions long known to be compatible with Raman spectroscopy (low temperatures, high pressures, high magnetic fields). The method is inspired by, and will complement, existing rotational-anisotropy optical probes, such as nonlinear harmonic generation and pump-probe polarimetry[39–43], by providing spectroscopic and polarimetric information without stringent requirements on the parent symmetry; in particular, unlike rotational anisotropy second-harmonic generation — which is primarily sensitive to inversion-symmetry breaking — RA-Raman imposes no such constraint.

(2) Chiral collective excitations/quasiparticles. RA-Raman also opens a route to detect lattice, spin, and electronic excitations with unconventional topological character, including chiral phonons, chiral magnons, and magnon-phonon hybrids in structurally chiral systems or in materials with broken time-reversal symmetry[44–48]. At its core, RA-Raman unites spectroscopy (sensitive to the energy of the excitation) with polarimetry (sensitive to the structure of the eigenmode/wavefunction): conventional spectroscopic tools excel at the former but offer limited access to the latter. In particular, antisymmetric Raman tensor components indicative of axial character can be accessed with circularly polarized input-output configurations and revealed via Raman optical activity signatures[49,50]. Leveraging RA-Raman's ability to resolve the full Raman-tensor structure, this approach can be extended broadly to characterize eigenmode symmetry and drive the discovery of topologically nontrivial excitations.

**Supporting Information**

Supporting Information is available from a separate file provided by the authors.

**Acknowledgements**

X.L. acknowledges support from the Singapore National Research Foundation under award no. NRF-NRFF16-2024-0008.

**Conflict of Interest**

The authors declare no conflict of interest.

**Data Availability Statement**

Full access to data supporting the plots within this paper will be deposited in public structured data depository. Data supporting other findings of this study are available from the corresponding author upon reasonable request.